\documentclass[10pt]{article}
\usepackage{jheppub}
\usepackage{enumerate}
\usepackage{bm}
\usepackage{bbm}
\usepackage[usenames,dvipsnames]{xcolor}

\usepackage{amssymb,amsmath}
\usepackage{graphicx}
\usepackage{overpic}
\usepackage{color}
\usepackage[colorlinks=true]{hyperref}
\usepackage{enumitem}
\usepackage{braket}
\usepackage{tensor}

\def\ba{\begin{eqnarray}}
\def\ea{\end{eqnarray}}
\newcommand{\mpl}{M_{\rm{Pl}}}

\newcommand{\D}{\rm{D}}
\newcommand{\tC}{\tilde{C}}

\newcommand{\K}{\mathcal{K}}

\def\({\left(}
\def\){\right)}

\begin{document}

\title{Black Hole Gravitational Waves in the Effective Field Theory of Gravity}

\author[a,b]{Claudia de Rham}
\author[c]{J\'er\'emie Francfort}
\author[a]{\& Jun Zhang}
\affiliation[a]{Theoretical Physics, Blackett Laboratory, Imperial College, London, SW7 2AZ, UK}
\affiliation[b]{CERCA, Department of Physics, Case Western Reserve University, 10900 Euclid Ave, Cleveland, OH 44106, USA}
\affiliation[c]{D\'epartement de Physique Th\'eorique, Universit\'e de Gen\`eve, 24 quai Ansermet, CH–1211 Gen\`eve 4, Switzerland}

\emailAdd{c.de-rham@imperial.ac.uk}
\emailAdd{jeremie.francfort@unige.ch}
\emailAdd{jun.zhang@imperial.ac.uk}

\abstract{We investigate the propagation of gravitational waves on a black hole background within the low--energy effective field theory of gravity, where effects from heavy fields are captured by higher--dimensional curvature operators. Depending on the spin of the particles integrated out, the speed of gravitational waves at low--energy can be either superluminal or subluminal as compared to the causal structure observed by other species. Interestingly however,  gravitational waves are always exactly luminal at the black hole horizon, implying that the horizon is identically defined for all species. We further compute the corrections on quasinormal frequencies caused by the higher dimensional curvature operators and highlight the corrections arising from the low--energy effective field. }

\maketitle

\section{Introduction}

The detection of Gravitational Waves (GWs) opens up a brand new window of opportunity to test gravity. The observation of GW170817 \cite{TheLIGOScientific:2017qsa} together with its gamma--ray counterpart GRB170817A \cite{Goldstein:2017mmi} constrains the speed difference between GWs and photons propagating on a cosmological background down to $10^{-15}$ \cite{Monitor:2017mdv}. In this new era of GW astronomy, it has become more important than ever to understand how GWs propagate especially in the strong gravity regime and get a handle on the types of corrections that are expected to arise in the effective field theory of gravity.  \\

Lorentz invariance dictates that in the vacuum any massless particle propagates at the speed of light, but in media that (spontaneously) break Lorentz invariance, we are used to expect a frequency--dependent deviation from luminal propagation, as is well known for light propagating through glass or water.  This effect emerges naturally from the interactions between light and the medium it propagates through. The  speed of photons can also be modified in a curved background due to loop corrections from charged particles (e.g. electrons). At energy scales well below the charged particle mass,  the low--energy effective field theory contains operators that can lead to a superluminal group and phase velocity on certain backgrounds \cite{Lafrance:1994in, Drummond:1979pp}. Yet this low--energy superluminal group and phase velocity is not in conflict with causality as discussed in \cite{Shore:1995fz,Shore:2000bs,Hollowood:2009qz,Hollowood:2015elj,Hollowood:2007kt,Hollowood:2007ku,Hollowood:2008kq,Hollowood:2010bd,Hollowood:2010xh,Hollowood:2011yh,Hollowood:2012as,Goon:2016une,deRham:2020zyh}.\\

By analogy, the same is expected for GWs. When accounting for the interactions between gravity and the other heavy fields one would expect the speed of GWs to naturally depart from unity at low--energy in backgrounds that spontaneously break Lorentz invariance, while recovering a luminal speed at high energy \cite{deRham:2018red}.  \\

In order to remain general and agnostic on the precise high--energy completion of gravity (i.e. on the precise spectrum of the heavy fields considered), we work here within the low--energy effective field theory (EFT) of gravity framework, where the classical and quantum effects of heavy fields is captured by the inclusion of higher--dimensional curvature operators \cite{Donoghue:1994dn,Donoghue:1995cz,Burgess:2003jk,Donoghue:2012zc}.
Indeed, we expect these operators to naturally arise from an arbitrary underlying UV complete gravity theory, such as string theory \cite{Metsaev:1986yb,Gross:1986iv,Gruzinov:2006ie}, although we do not need to commit to any particular realization in what follows.
 Within this low--energy EFT of gravity, it was shown in  \cite{deRham:2019ctd} that GWs propagating in a FLRW background do not generically propagate exactly luminally at low--energy. In this work, we push this investigation further by analyzing  the speed of GWs propagating on a Schwarzschild--like background. This will represent an interesting situation where the GWs are propagating in the vacuum but the presence of a black hole spontaneously breaks Poincar\'e invariance. This implies that the low--energy speed of GWs can (and indeed does) differ from the speed of other massless minimally--coupled particles.   Since the speed of various species is not invariant under change of frames, we qualify our statement and make the impact on the causal structure manifest by working in Jordan frame, where all the matter fields (including light) are minimally coupled to gravity, ensuring that electromagnetic waves travel at a luminal speed with respect to the background metric. In this frame, we consider the low--energy EFT of gravity by including the local and covariant higher order curvature operators present in the low--energy EFT.  These can emerge from weakly coupled UV completions after integrating out fields of higher--spin at tree level, or can emerge from integrating out loops of particles of all spins, including standard model particles (see Ref.~\cite{deRham:2019ctd} for a detailed discussion). Once again, for the most part of this work, we shall remain agnostic on the precise realization. \\

If treated non-perturbatively the higher--dimensional curvature operators may lead to interesting features as  pointed out for instance in \cite{Cayuso:2020lca}. In what follows we shall however take an EFT approach considering the higher--dimensional operators to represent only the leading contributions in an infinite low-energy expansion.
Since the low--energy EFT is only meaningful at energy scales well--below the cutoff, all higher--dimensional curvature operators should be understood as being treated perturbatively and this is indeed the approach we shall take in what follows. Working perturbatively implies that the dimension--4 (curvature--squared) operators do not affect the evolution of GWs when we restrict ourselves to a background perturbatively connected to the GR Schwarzschild background\footnote{In principle, there could be other branches of solutions in a theory with higher dimension operators, but those solutions rely on exciting the higher dimensional operators beyond the regime of validity of the low--energy EFT.}.
To determine the leading order corrections on the propagation of GWs, we therefore have to consider curvature dimension--6 (curvature--cubed) operators.
 Including the perturbative contribution from these operators, we extract the effective metric seen by the metric perturbations and identify the speed of GWs. We find that the speed indeed deviates from the speed of photons in general. As expected, the deviation caused by the higher--dimensional operators is highly suppressed. However any departure from unity is significant in itself, as it reflects the causal structure of the theory. In particular, the deviation vanishes as one approaches the horizon. This remarkable feature implies that while GWs and photons see a different causal structure almost everywhere, they still experience the horizon at the precise same location. We argue that this has to always be the case. For completeness we also compute the corrections on the quasinormal frequencies of the black holes in the EFT of gravity. As expected the corrections from the EFT operators are extremely suppressed and determined in terms of only two of the dimension--6 EFT operators.\\

The rest of this paper is organised as follows. In Section~\ref{sec:BH}, we introduce the low--energy EFT of gravity, including the dimension--6 operators. We study their perturbative effects on the black hole solution and derive the modified Regge--Wheeler--Zerilli equations for the metric perturbations. This allows us to investigate the speed of GWs and the causal structure in Section~\ref{sec:GW}. A potential connection with the horizon theorem is also discussed. The corrections on the black hole quasinormal frequencies are computed in Section~\ref{sec:QNM}. Section~\ref{sec:dis} is devoted to discussions and outlook. Technical details and some expressions are given in the Appendixes. We work with the $(-,+,+,+)$ signature, and in units where $\hbar=c=1$.

\section{Black Holes in the Low--Energy Effective Field Theory of Gravity}\label{sec:BH}

\subsection{EFT of gravity}

We consider the low--energy EFT of gravity including  curvature operators up to dimension--6. The Lagrangian of the theory is given by \cite{Metsaev:1986yb}
\ba
\label{eq:Ltotal}
\mathcal{L} = \sqrt{-g} \frac{\mpl^2}{2} R + \mathcal{L}_{\rm D4} + \mathcal{L}_{\rm D6}+\mathcal{L}_{\text{light matter fields}}(g, \psi)+\mathcal{O}\left(\frac{{\rm Riemann}^4}{M^4}\right),
\ea
where $\psi$ designates symbolically all the light fields (including the photon) that are explicitly included within the low--energy EFT. The dynamics of these fields will not be relevant for this study as we shall be interested in vacuum solutions. The higher--dimensional operators are given by
\ba\label{eq:l2}
\mathcal{L}_{\rm D4} &=& \sqrt{-g} \left[c_{R^2} R^2 + c_{W^2} W_{\mu\nu\alpha\beta}^2 + c_{{\rm GB}}R^2_{\rm GB} \right],
\ea
and
\ba
\label{eq:l3}
\mathcal{L}_{\rm D6}=\frac{1}{M^2}\sqrt{-g}&&\Big[ d_1 R\Box R + d_2 R_{\mu \nu} \Box R^{\mu \nu} +d_3 R^3 + d_4 R R_{\mu \nu}^2   \nonumber \\
&&+ d_5 R R_{\mu \nu \alpha\beta}^2 + d_6 R_{\mu \nu}^3 + d_7 R^{\mu \nu} R^{\alpha \beta}R_{\mu\nu\alpha\beta} + d_8 R^{\mu\nu} R_{\mu\alpha \beta \gamma} {R_{\nu}}^{\alpha \beta \gamma} \nonumber \\
&&+  d_9 \tensor{R}{_\mu_\nu^\alpha^\beta} \tensor{R}{_\alpha_\beta^\gamma^\sigma} \tensor{R}{_\gamma_\sigma^\mu^\nu}+ d_{10}\tensor{R}{_\mu^\alpha_\nu^\beta}\tensor{R}{_\alpha^\gamma_\beta^\sigma} \tensor{R}{_\gamma^\mu_\sigma^\nu}\Big],
\ea
where $R^2_{\rm GB} = R_{\mu \nu \alpha\beta}^2 - 4 R_{\mu \nu}^2 + R^2$ is the Gauss--Bonnet term, and $W_{\mu\nu\alpha\beta}$ is the Weyl tensor. In four dimensions, the Gauss--Bonnet term is topological, which allows us to rewrite the dimension--4 curvature operator Lagrangian as
\ba
\label{eq:D4}
\mathcal{L}_{\rm D4} =\sqrt{-g} \left[ c_1 R^2 + c_2 R_{\mu\nu}R^{\mu\nu} \right],
\ea
with
$$
c_1 = c_{R^2}- \frac{2}{3} c_{W^2}, \quad  c_2 = 2c_{W^2}.
$$
Since we shall be  interested in vacuum solutions with $R_{\mu\nu}=0$, it is therefore clear that the dimension--4 operators cannot lead to any leading order\footnote{At second order in perturbations, the dimension--4 operators can lead to non--trivial effects, however those will be suppressed by a factor of $M^2/\mpl^2$ as compared to the leading order effects from dimension--6 operators.} correction, neither in the background solution nor in the propagation of GWs. In the rest of this manuscript we shall therefore focus our interest on the dimension--6 operators. Moreover, in four dimensions, the  operator governed by $d_{10}$ can be written as a combination of other dimension--6 operators \cite{Cano:2019ore}, and can therefore be removed\footnote{We thank Pablo A. Cano for pointing this out.}. To compare the Lagrangian~\eqref{eq:l3} with the EFT arising from integrating out a heavy field \cite{Avramidi:1990je,Avramidi:1986mj}, we shall however keep $d_{10}$ manifest. \\

If we consider the Lagrangian \eqref{eq:Ltotal} as the low--energy EFT generated arising from integrating out some heavy fields, we expect the dimension--6 curvature operators to be suppressed by the mass $M$ of the lightest of the massive fields being integrated out (i.e. the lightest of all the fields that are not explicitly included in $\mathcal{L}_{\text{light matter fields}}(g, \psi)$).\\

It is known that some of the higher--dimensional curvature operators can be removed by field redefinition, however performing such field redefinition will introduce interactions in the matter sector \cite{deRham:2019ctd}, and hence alter the photon speed (see also Ref.~\cite{Burrage:2016myt} where a similar point in a slightly different context has been made). To make the impacts on the causal structure manifest, we shall therefore stick to the  frame in which the speed of photons is unity. Of course one could start with the field--redefined Lagrangian that includes fewer operators and revert back to the original frame at the end. However this method does not prove optimal at the computational level as it will introduce subtleties in gauge fixing when reverting back to the original frame, see Appendix~\ref{app:redef} for more details. Note however that for the particular vacuum solution we are interested in, not all operators present in \eqref{eq:l3} contribute physically. Actually, as motivated in Appendix~\ref{app:redef}, only the coefficients $8d_5 +2d_{8} - 3 d_{10} $ and $2d_9+ d_{10}$ contribute to the background solution and the dynamics of GWs.

\subsection{Dimension--6 vs dimension--8 operators}
\label{sec:dim6_8}

At this stage we should note that black hole perturbations in the EFT of gravity were already previously considered in \cite{Endlich:2017tqa,Cardoso:2018ptl}. The emphasis of \cite{Cardoso:2018ptl} was primarily the study of quasinormal modes while we shall here be primarily interested in the speed of GWs and as explained previously, such effects are not invariant under field redefinitions. It is therefore relevant for our analysis to maintain operators in the EFT even if those could a priori  be removed via field redefinitions.  \\

Moreover, in the EFT considered in \cite{Endlich:2017tqa}, the focus was drawn on dimension--8 operators (for instance Riemann$^4$--types of operators). Assuming a weakly coupled UV completion, the dimension--6 types of operators can only be present upon integrating higher spin particles, whose mass is directly related to the scale $M$ of the EFT. Within such a completion, the absence of observable effects from higher--spin particles therefore puts a constraint on the scale $M$. The same argument goes through for dimension--8 operators and assuming a weakly coupled UV completion, the dimension--8 operators should themselves also be further suppressed. In fact unless one assumes the existence of very specific tuning, one would always expect dimension--6 operators to dominate over the dimension--8 and higher operators (the only reason the dimension--4 operators do not dominate in this setup is an accident of being in four dimensions and in the vacuum). Moreover, in this study we shall not commit to a weakly--coupled UV completion and the operators considered in \eqref{eq:Ltotal} may come either from integrating out higher spin particles at tree--level or from integrating out loops of particles of any spin \cite{deRham:2019ctd}, generic completions can indeed have various effects on the low-energy EFT \cite{Alberte:2020jsk}.  For these reasons we shall focus on dimension--6 operators in what follows. \\

Naturally, the size of the corrections we are studying is expected to be tiny at best but the question we are establishing is first whether in principle the low--energy speed of GWs could ever deviate ever so slightly from the ``speed of light" as dictated from the background metric and second
to determine the location of the horizon as seen by low-frequency GWs.
As we shall argue in Section~\ref{sec:GW}, validity of the EFT at the horizon dictates that the location of the horizon should always remain precisely the same for any species present in the low-energy EFT and this is indeed what we observe in our framework.

\subsection{Spherically symmetric black hole solutions}
\label{sec:EFTBackground}
Since we shall be interested in static and  spherically symmetric solutions, we make the Ansatz,
\ba\label{eq:ansatz}
\mathrm{d}s^2 = \bar{g}_{\mu\nu} \mathrm{d} x^{\mu} \mathrm{d}x^{\nu}= - A(r) \mathrm{d}t^2 + \frac{1}{B(r)} \mathrm{d}r^2 + C(r) r^2 \left(\mathrm{d}\theta^2 + \sin^2\theta\, \mathrm{d}\phi^2\right).
\ea
Substituting the Ansatz into Lagrangian \eqref{eq:Ltotal} and varying the Lagrangian with respect to $A$, $B$ and $C$ yield equations ${\cal E}_A$, ${\cal E}_B$ and ${\cal E}_C$. We shall look for the slight deviations from the Schwarzschild geometry caused by the higher--dimension curvature operators. As the dimension--4 operators do not contribute to the Ricci flat solutions, the leading corrections are caused by the dimension--6 operators. Therefore, deviations from the Schwarzschild geometry should be suppressed by a dimensionless small parameter
\ba\label{eq:epsilon}
\epsilon = \frac{1}{ M^2 \mpl^2 r_g^4}\,,
\ea
with $r_g$ being the Schwarzschild radius of the GR black holes. Choosing the gauge so that
$C(r) =1$ and solving $A$ and $B$ to the first order in $\epsilon$, we find
\ba\label{eq:ABsol}
A(r) &=&  1 - \frac{r_g}{r} + \epsilon \left[ a_6 \left(\frac{r_g}{r}\right)^6 + a_7 \left(\frac{r_g}{r}\right)^7 \right], \\
B(r) &=&  1 - \frac{r_g}{r} + \epsilon \left[ b_6 \left(\frac{r_g}{r}\right)^6 + b_7 \left(\frac{r_g}{r}\right)^7 \right],
\ea
where
\ba\label{eq:cis}
&&a_6 =  - 6\, d_{58} + 9\, d_{10} , \quad a_7 = \frac{1}{2}\left( 18\, d_{58} + 20\, d_9 - 17\, d_{10}\right), \\
&&b_6 =  36\, d_{58}+ 108\, d_9 , \quad  b_7 = \frac{1}{2}\left(- 66\, d_{58} - 196\, d_9 + d_{10}\right),
\ea
with $d_{58} \equiv 4 d_5 + d_8$.
Since the higher--dimensional operators cannot lead to any physical singularity within the region of validity of the EFT,
to this order $A$ and $B$  must vanish simultaneously $A(r=r_H)  = B(r=r_H)  = 0  +  {\cal O}(\epsilon^2)$ at the same point, defining the location $r_H$ of the perturbed horizon in the EFT. This is indeed the case and the horizon of the background metric $\bar{g}_{\mu\nu}$ is defined as
\ba\label{eq:rH}
r_H \equiv r_g -\epsilon\left(3\, d_{58} + 10\, d_9 +\frac12\, d_{10}\right)r_g\,.
\ea
As emphasized in Appendix~\ref{app:redef} one can check that only the operators $d_{9}$, $d_{10}$ and $d_{58}\equiv 4d_5 + d_8$ enter the background vacuum solution. A similar study on black holes in an EFT of gravity in the presence of higher dimensional operators was performed in \cite{Cano:2019ore,Cano:2020cao}, where the coefficients of the higher dimensional-operators were considered to be dynamical and controlled by scalars. The emphasis of our analysis is however different as we shall mainly be interested in the dynamics of GWs on this geometry.

\subsection{Black hole perturbations}
We now consider metric perturbations about the previous background solution.  We first start with the covariant equations of motion, which can be written as
\ba\label{eq:coveq}
\frac{\mpl^2}{2} G_{\mu\nu} +\frac{1}{M^2} {\cal E}_{\mu\nu} = 0.
\ea
At zeroth order in $\epsilon$ (i.e. for pure GR in the vacuum), the vacuum Ricci flat solutions have
\ba
R_{\mu\nu}= \delta R_{\mu\nu} = 0,
\ea
where $\delta R_{\mu\nu}$ is the perturbations of Ricci tensor caused by the metric perturbations. It follows that at leading order in the EFT corrections, any term in the Lagrangian that is quadratic in $R_{\mu\nu}$ will not  affect the evolution of the metric perturbations. The only relevant contributions in ${\cal E}_{\mu\nu}$ are therefore
\ba
\label{eq:EOMR3}
{\cal E}_{\mu\nu} &=&
d_5 \left(- \nabla_\mu \nabla_\nu C^\alpha_{\, \alpha} + g_{\mu\nu} \Box C^\alpha_{\, \alpha} \right) \\
&+&
\frac{d_8}{2} \left(\Box C_{\mu\nu} - \nabla^\alpha \nabla_\mu C_{\nu \alpha} - \nabla^\alpha \nabla_\nu C_{\mu \alpha} +  g_{\mu\nu} \nabla_\alpha \nabla_\beta C^{\alpha \beta} \right) \notag \\
&+&
d_9 \left(6 \nabla^\alpha \nabla^\beta C_{\mu\alpha\nu\beta} -\frac12 g_{\mu\nu} R^{\alpha\beta\gamma\sigma}C_{\alpha\beta\gamma\sigma} + 3 \tensor{R}{_\mu^\alpha^\beta^\gamma} C_{\nu \alpha \beta \gamma}\right) \notag \\
&+&
d_{10} \left( \frac32 \nabla^\alpha \nabla^\beta \tC_{\mu\alpha\nu\beta}+ \frac32 \nabla^\alpha \nabla^\beta \tC_{\nu\alpha\mu\beta}-  \frac32\nabla^\alpha \nabla^\beta \tC_{\mu\beta\alpha\nu}- \frac32 \nabla^\alpha \nabla^\beta \tC_{\nu\beta\alpha\mu}  \right. \notag \\
&&\left. - \frac12 g_{\mu\nu} R^{\alpha\beta\gamma\sigma}\tC_{\alpha\beta\gamma\sigma} + 3 \tensor{R}{_\mu^\alpha^\beta^\gamma} \tC_{\nu \alpha \beta \gamma}\right),\notag
\ea
where we have defined the two contractions of the Riemann tensor,
\ba
 \tensor{C}{_\mu_\nu^\gamma^\sigma} \equiv  {R_{\mu\nu}}^{\alpha\beta} {R_{\alpha\beta}}^{\gamma\sigma}, \quad C_{\mu\nu} = \tensor{C}{_\mu_\alpha_\nu^\alpha}, \quad {\rm and} \quad \tensor{\tC}{_\mu^\gamma_\nu^\sigma} \equiv \tensor{R}{_\mu^\alpha_\nu^\beta} \tensor{R}{_\alpha^\gamma_\beta^\sigma}.
\ea
One can show that the indices of $\tensor{C}{_\mu_\nu^\gamma^\sigma} $ are Riemann symmetric, and $\tC_{\mu\nu \gamma\sigma} = \tC_{\gamma\sigma\mu\nu}= \tC_{\nu\mu \sigma\gamma}$.

\subsubsection{Metric perturbations}

We decompose the metric perturbations into their odd $(\rm o)$ and even $(\rm e)$ contributions, $h_{\mu\nu}=h_{\mu\nu}^{\rm o} + h_{\mu\nu}^{\rm e}$ based on their behaviours under parity transformations $(\theta, \phi) \rightarrow (\pi-\theta, \pi+\phi)$. Indeed,  the spherical symmetry of the background allows us to restrict ourselves to axisymmetric modes of perturbations without any loss of generality. Non--axisymmetric modes, i.e., perturbations with an $e^{i m \phi}$ dependence, can be deduced from modes of axisymmetric perturbations with $m=0$ by suitable rotations. We shall work in the Regge--Wheeler gauge \cite{Regge:1957td}, in which
\ba\label{eq:habodd}
h_{\mu\nu}^{\text{o}}= e^{-i \omega t} \begin{pmatrix}
0 &0  & 0 & h_0   \\
0 & 0  & 0 & h_1 \\
0 & 0 & 0 & 0 \\
 h_0 &  h_1 &0 &0
\end{pmatrix}  \sin \theta \, Y'_{\ell}(\theta) ,
\ea
and
\begin{equation}
h_{\mu\nu}^{\text{e}} = e^{-i \omega t} \begin{pmatrix}
A H_0 & H_1  & 0&0 \\
H_1 & H_2/B  & 0& 0 \\
0 & 0 &r^2 \mathcal{K}  & 0\\
0 & 0 & 0 & r^2 \sin^2\theta \, \mathcal{K}
\end{pmatrix} Y_{\ell}(\theta) \, , \label{eq:evenhab}
\end{equation}
where $Y_{\ell}(\theta) = Y_{\ell 0} (\theta, \phi)$ are the spherical harmonics with $m=0$, and a prime on $Y_\ell$ denotes the derivative with respect to the angle $\theta$. All the functions $h_0$, $h_1$, $H_0$, $H_1$, $H_2$,  and $\mathcal{K}$ are functions of $r$.

\subsubsection{Master equations}
\label{sec:masterEqs}

\noindent $\bullet$ {\bf Regge--Wheeler and Zerilli equations in GR:}
The symmetry of the background already allows us to restrict ourselves to the six metric perturbations $h_{0,1}$, $H_{0,1,2}$ and  $\mathcal{K}$ introduced in \eqref{eq:habodd} and \eqref{eq:evenhab}, but we know that only two degrees of freedom are present in GR which in this case should manifest themselves as one odd and one even mode. The dynamics of each one of these physical modes $\Psi^{\rm o/e}$ is determined by the Regge--Wheeler and Zerilli equations,
\ba\label{eq:GRmaster}
\frac{{\rm d}^2\Psi_{\rm GR}^{\text{o/e}}}{{\rm d}r_*^2}+ \left[\omega_0^2  -\left(1-\frac{r_g}{r}\right)V_{\rm GR}^{\text{o/e}}\right]\Psi_{\rm GR}^{\text{o/e}} = 0,
\ea
with
\ba
&&V_{\rm GR}^{\text{o}} = \frac{J}{r^2} - \frac{3r_g}{r^3}, \\
&&V_{\rm GR}^{\text{e}} =\frac{J(J-2)^2 r^3+3 (J-2)^2 r^2 r_g+9 (J-2) r r_g^2+9 r_g^3}{r^3 [(J-2) r+3 r_g]^2},
\ea
where we have defined $J \equiv \ell(\ell+1)$, the tortoise coordinate ${\rm d}r_*= {\rm d}r/(1-r_g/r) $,  and the two master variables
\ba
&&\Psi_{\rm GR}^{\text{o}} \equiv \frac{i (r- r_g )\, h_1}{r^2 \omega_0}, \\
&&\Psi_{\rm GR}^{\text{e}} \equiv \frac{1}{(J-2)r + 3 r_g}\left[-r^2 \K + \frac{i (r-r_g)H_1}{ \omega_0}\right].
\ea
The other components of $h_{\mu\nu}$ are uniquely determined (constrained) in terms of  $\Psi^{\text{o/e}}$. \\

\noindent $\bullet$ {\bf Leading order corrections from the dimension--6 EFT operators:}
Including the leading order corrections from the dimension--6 operators leads to higher--derivative equations of motion. We emphasize however that there is no sense in which these higher derivatives should ever be associated with additional Ostrogradsky ghost degrees of freedom within the regime of validity of the EFT. Indeed the mass of those would--be ghosts would always be at or above the cutoff of the low--energy EFT\footnote{The emergence of the ghost--like instability only arises from exciting modes  which lie beyond the regime of validity of the EFT \cite{deRham:2014fha}. Moreover, these would--be ghosts should not be identified with the heavy degrees of freedom that have been integrated out \cite{deRham:2018dqm}, rather their existence is a simple manifestation of applying an EFT beyond its regime of validity.}. Within the regime of validity of the EFT, the effects from the dimension--6 operators ought to be treated perturbatively and any higher order derivative should be removed using the lower order equations of motion (see Ref.~\cite{deRham:2019ctd} for a generic prescription). Doing so to leading order results in the following two second order differential equations,
\ba\label{eq:master}
\frac{{\rm d}^2\Psi^{\text{o/e}}}{{\rm d}r_*^2}+  \frac{ \omega^2}{c_s^2} \Psi^{\text{o/e}} -\sqrt{AB}\left[V_{\rm GR}^{\text{o/e}} + \epsilon\,  V^{\text{o/e}} \right]\Psi^{\text{o/e}} = 0
\ea
with
\begin{eqnarray}
\label{eq:cs}
c_s^2 = 1 - \epsilon \, \Delta c +\mathcal{O}(\epsilon^2),\quad {\rm and} \quad \Delta c =144 (2 d_9 + d_{10}) \frac{(r-r_g)r_g^5}{r^6}\,.
\end{eqnarray}
The profile of the low--energy radial speed $c_s$ is depicted in Fig.~\ref{fig:cs3} and discussed in detail in Section~\ref{sec:GW}.

Note that the tortoise coordinates are now defined by ${\rm d}r/{\rm d}r_* = \sqrt{AB}$ and the master variables are expressed as
\ba\label{eq:variable}
&&\Psi^{\rm o} = \frac{i \sqrt{A B}\, h_1}{r \omega} \bigg[1+ \epsilon f_{h_1} \bigg], \\
&&\Psi^{\rm e} = \frac{1}{(J-2)r + 3 r_g}\left[-r^2 \K (1+\epsilon f_\K)+ \frac{i \sqrt{AB}\,rH_1}{ \omega}(1+\epsilon f_{H_1})\right].
\ea
The explicit expression of $V^{\rm o/e}$, $f_{h_1}$, $f_\K$ and $f_{H_1}$ can be found in Appendix~\ref{app:exp}. A similar analysis on black holes in EFTs with dimension--8 curvature operators can be found in \cite{Cardoso:2018ptl}. See Section~\ref{sec:dim6_8} for a discussion of why dimension--6 are considered in this work as opposed to dimension--8 operators.

\section{Low--Energy Speed of Gravitational Waves}\label{sec:GW}

While the speed of GWs is exactly $c_s\equiv 1$ in GR (i.e. in the absence of higher--order curvature operators), we see from \eqref{eq:cs} that on the background of the black hole (that spontaneously breaks Poincar\'e--invariance), the speed of GWs departs ever so slightly from unity once the irrelevant operators from the EFT of gravity are taken perturbatively into account. Note that this low--energy EFT breaks down at the scale $M$ (or even lower), and at sufficiently high energy, GWs would recover exact luminality.

\begin{figure}[h]
\centering
\includegraphics[width=0.6\textwidth]{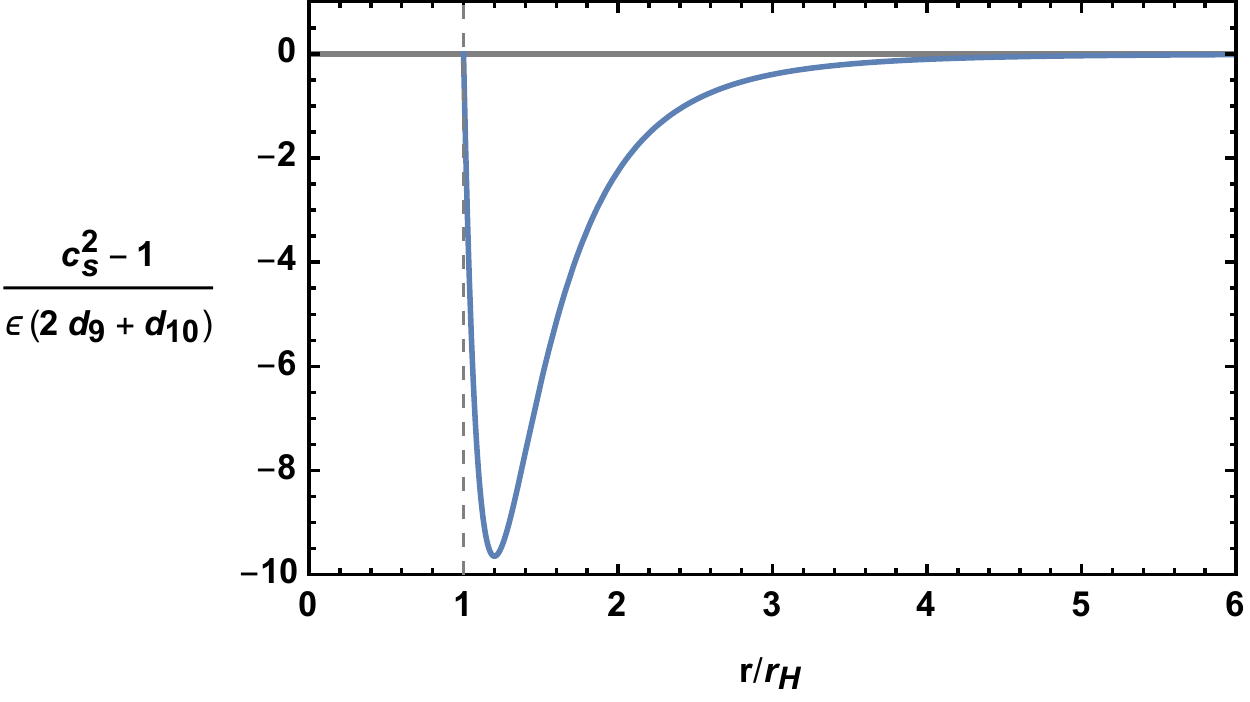}
\caption{Deviation of the low--energy radial speed $c_s$ from unity. The deviation is maximised at $r=6 r_H/5$, and vanishes at the horizon $r=r_H$ as well as asymptotic infinity $r\rightarrow \infty$. The coefficient $(2 d_9 + d_{10})$ can be a priori of either sign in the low--energy EFT.  For instance if we see this EFT as arising from integrating out a heavy field, the coefficient $(2 d_9 + d_{10})$ is positive (resp. negative) for a particle of spin--$0$ or $1$ (resp. spin $1/2$) \cite{Avramidi:1990je, Avramidi:1986mj}. GWs would then be subluminal if the lightest particle being integrated out is a  scalar or a vector, and  superluminal if it is a fermion.} \label{fig:cs3}
\end{figure}

Within the frame we are working in, the speed of photons and other massless particles remains unity, (assuming there is no direct coupling between these massless particles and the heavy particles that have been integrated out). Deviation of $c_s^2$ from unity is shown in Fig.~\ref{fig:cs3}, which is non--zero everywhere outside the black hole but remarkably vanishes at the horizon $r=r_H$.

The sign of $c_s^2-1$ and therefore whether GWs are expected to be ever so slightly sub or super luminal depends on the precise UV completion. For instance if the EFT we are considering was arising from integrating out particles of spin--$0$, $1/2$ and $1$, the precise value of the coefficients $d_i$ would be dictated by the spins of these particles. as derived in \cite{Avramidi:1990je, Avramidi:1986mj}. In this case, the coefficient $(2 d_9 + d_{10})$ is positive for scalars and vectors and negative for fermions. In other words, depending on the precise field content (or on the spin of the lightest massive particle that has been integrated out),  GWs may turn out to be ever so slightly subluminal or superluminal.

As mentioned in the introduction, superluminal low--energy group and phase velocities do not necessarily lead to violation of causality, see Ref.~\cite{deRham:2020zyh}. Nevertheless these types of arguments have been used in the past to segregate between various types of EFTs. Applying these types of arguments to the situation at hand, one would conclude that  neutrinos cannot be the lightest massive particles and one would conjecture the existence of lighter particles of different spin. However we would caution against applying these types of arguments when it comes to the EFT of gravity where the size of the corrections is so small that no violation of macrocausality can even occur \cite{deRham:2020zyh}.

To extract the effective metric seen by GWs (in the frame where other matter fields see the background metric) let us consider a scalar $\Phi$ propagating on an effective metric $Z_{\mu\nu}$
\ba\label{eq:scalara}
Z^{\mu\nu} \D_\mu \D_\nu \Phi + {\rm U} \Phi = 0,
\ea
where ${\rm U}$ is an effective potential, $\D_\mu$ represents the covariant derivative  with respect to $Z_{\mu\nu}$, and $Z^{\mu\alpha}Z_{\alpha\nu} = \delta^{\mu}_{\nu}$ with
\ba
Z_{\mu\nu} =
\begin{pmatrix}
	-Z_t(r) &0  &0  &0 \\
	0 & Z_r^{-1}(r) &0  &0 \\
	0 & 0 & Z_{\Omega}(r)r^{2} &0 \\
	0 & 0 &0  & Z_{\Omega}(r)r^{2} \sin^{2} \theta
\end{pmatrix}.
\ea
Substituting $\Phi = e^{-i \omega t} \Psi(r) Y_\ell (\theta) / r^2$ into Eq.~\eqref{eq:scalara} yields
\ba\label{eq:scalarb}
\Psi'' + \left[  \frac{(Z_r Z_t)'}{2 Z_r Z_t} + \frac{Z_\Omega'}{Z_\Omega}  \right]\Psi' + \frac{\omega^2}{Z_t Z_r} \Psi - \frac{J}{r^2 Z_{\Omega}Z_r}\Psi + \frac{{\rm U}}{r^2 Z_r}\Psi= 0,
\ea
where a prime denotes the derivative with respect to $r$. On the other hand, Eq.~\eqref{eq:master} can be written in the form
\ba\label{eq:form}
\Psi'' + \frac{(AB)'}{2AB}\Psi' + \frac{\omega^2}{c_s^2 A B} \Psi- \frac{J}{r^2 \sqrt{AB}}\Psi + V\Psi = 0,
\ea
where we have neglected the superscription ``o/e" for simplicity. Comparing Eq.~\eqref{eq:form} with Eq.~\eqref{eq:scalarb}, we can read off
\ba\label{eq:eqc}
\frac{(Z_r Z_t)'}{2 Z_r Z_t} + \frac{ Z_\Omega'}{ Z_\Omega}  = \frac{(AB)'}{2AB}\,, \quad Z_t Z_r = c_s^2AB \quad {\rm and} \quad Z_\Omega Z_r = \sqrt{AB}.
\ea
Including the leading order corrections from the EFT of gravity, we may now express $c_s^2 = 1- \epsilon \, \Delta c$, with $\Delta c$ given in \eqref{eq:cs}
and Eq.~\eqref{eq:eqc} then implies
\ba
Z_t =Z_r=\sqrt{AB} \left(1- \frac12  \epsilon\,  \Delta c\right), \quad \text{and} \quad  Z_\Omega = 1+  \frac 12 \epsilon\,  \Delta c\,.
\ea
To leading order in the EFT, this corresponds to\footnote{The angular part of the effective metric can be expressed in the usual way by redefining the radial coordinate in a way that does not affect the causal structure.}
\ba
Z_t = Z_r =
 1-\frac{r_g}{r}
 + \epsilon \frac{r_g^5}{2r^5}\Bigg[6 d_{58}  \left(5 \frac{r_g}{r}-4 \frac{r_g^2}{r^2}\right)
 -4 d_9 \left(72 -171 \frac{ r_g}{r}+94 \frac{r_g^2}{r^2}\right)\\ -d_{10} \left(144 -297 \frac{r_g}{r}+152 \frac{r_g^2}{r^2}\right)\Bigg]\nonumber\,.
\ea
At at $r=r_H$, we find that $Z_t(r_H) = Z_r(r_H)= 0 + {\cal O}(\epsilon^2)$, and $Z_\Omega(r_H)=1$, which is exactly the horizon seen by photons in this EFT.

\paragraph{Connection with the Horizon Theorem:}
Interestingly, the previous result shows that both GWs and photons see the same horizon at least to leading order in the EFT expansion, although their speeds and hence the causal structures are different near the black hole. A priori this result has only been shown here perturbatively to first order but it is  already non--trivial and should indeed have been expected to all order.\\

This result is closely connected to the more general proof derived within the context of the EFT of QED below the electron mass, known as the ``Horizon Theorem" \cite{Shore:1995fz}. In the context of the EFT of QED it was indeed shown that the irrelevant operators appearing in the EFT of QED necessarily have a vanishing effect at the horizon of any black hole irrespective of the precise structure and field/matter content, and therefore cannot affect the speed of light at the horizon of a black hole. It also follows that the black hole horizon remains a true horizon for light in this EFT.  Remarkably these results relied on very few assumptions and are generic to any stationary spacetimes \cite{Hawking:1973qla}. \\

In the context of QED, the Horizon Theorem was proven in \cite{Shore:1995fz} using properties of the Weyl tensor and its contractions with the four null-momentum of the photon derived in the Newman--Penrose tetrad basis \cite{Shore:2000bs,Hollowood:2009qz}.  While in principle the same type of formalism could be applied to the EFT of gravity, its implementation turns out to be quite subtle in practice. There are three complications that arise in the EFT of gravity as compared to that of QED.

The first one is that the EFT of gravity involve higher order equations of motion, which means that the dynamical equations only make sense perturbatively and one needs to make use of the lower--order equations of motion to make any progress. This first point makes the formalism slightly more subtle although in principle achievable.

A second, more problematic issue is that at the level of the perturbed equations of motion the Weyl tensor appears quadratically as opposed to linearly as in the case in \cite{Shore:1995fz} (see for instance Eq.~\eqref{eq:EOMR3} where the tensor $\tilde C$ plays a similar role to the Weyl--squared tensor). This implies that in order to make progress with this formalism one should generalize the relations derived for the complex scalars presented in Eq.~(2.14) of \cite{Shore:2000bs} or in Eq.~(7.1) of \cite{Hollowood:2009qz} to a new set of complex scalars involving contractions of two Weyl tensors and their derivatives. It would be interesting to establish and prove which subset of these complex scalars vanish at the horizon for stationary spacetimes, however such a proof would  not be straightforward and is beyond the scope of this current work. Such considerations are therefore saved for later studies.

Finally a third potential difference in the EFT of gravity could arise when considering non-vacuum backgrounds. In QED, the curved background can be caused by electromagnetically neutral matter, and photons decouple from gravitons at linear order (and hence decouple from matter perturbations). However this is typically not the case in the EFT of gravity, in which gravitons would always couple with the matter perturbations in non-vacuum backgrounds. It is likely that preserving the NEC and other consistency relations would be sufficient to make progress, but additional assumptions on the matter perturbations may be needed to draw conclusions in the EFT of gravity, even though one would ultimately expect the Horizon Theorem derived in \cite{Shore:1995fz} to apply to generic EFTs.\\

For lack of a more rigorous proof, we shall instead provide an intuitive EFT argument as to why one expects the black hole horizon to remain the true horizon for all the species in the EFT of gravity, including that of low-frequency GWs. First of all we recall that we expect the EFT of gravity to remain valid at the horizon. Indeed for macroscopic black holes we only expect the EFT to break down well--inside the black hole horizon.

With this expectation in mind,  we start with the metric seen by photons and other minimally coupled species as given in \eqref{eq:ansatz} with $C=1$ and with a horizon located at $r=r_H$.
Now imagine GWs saw another effective metric with an effective horizon located at a slightly different location $\tilde r_H=r_H+\epsilon\,  \delta r_H$. If this was the case, with $\delta r_H\ne 0$, then perturbatively close to $r_H$, the components $Z_{\mu\nu}$ of the effective metric of GWs would be finite at $r_H$,
\ba
\left.Z_t\right|_{r=r_H}=\left.Z_r\right|_{r=r_H}=-\epsilon \frac{\delta r_H}{r_H}+\mathcal{O}(\epsilon^2)\,.
\ea
One would then be able to compute a scalar invariant out of both metric $\bar g_{\mu \nu}$ and $Z_{\mu \nu}$ that diverges at the horizon. Indeed, denoting for instance by $W=W[\bar g]$ the Weyl tensor as computed with the metric $\bar g_{\mu \nu}$ and $\mathcal{W}=\mathcal{W}[Z]$ the Weyl tensor as seen by the low-energy gravitons, then out of these two slightly different versions of the Weyl tensor as seen by different species living in the same EFT, one would be able to construct a scalar invariant $\Upsilon$ defined as for instance
\ba
\Upsilon =W^{\mu\nu\alpha\beta}\mathcal{W}_{\mu\nu\alpha\beta}= - 2\frac{\delta r_H}{r_H^4}\frac{1}{r-r_H}\, \epsilon + \mathcal{O}\(\epsilon^2,\(1-\frac{r_H}{r}\)^0\)\,,
\ea
which would lead to a physical singularity at $r_H$, where the EFT is still valid and should have been under control. Since we are only computing physical quantities perturbatively in the EFT, all the corrections we are after are small and under control. Such a physical singularity can therefore never occur within the regime of validity of the EFT and we therefore conclude that we ought to have $\delta r_H=0$. In other words the horizon seen by gravitons ought to be the same as that of any other species present in this EFT.

\section{Quasinormal Modes}\label{sec:QNM}

For completeness, we end with a computation of the quasinormal modes. As already mentioned, we expect the size of the corrections from the EFT of gravity to be utterly negligible at best but stress that, within the regime of validity of the EFT of gravity, we always expect the effect to be dominated by the operators of lowest dimensionality. Since dimension--4 operators do not contribute at leading order in the vacuum, the dimension--6 operators are thus expected to lead to the ``dominant" corrections.\\

In what follows we shall be interested in the quasinormal frequency $\omega$ of black holes in the EFT of gravity \eqref{eq:Ltotal}. We start by denoting by  $\omega_0$ the quasinormal frequency of a GR black hole with Schwarzschild radius $r_g$, i.e., the quasinormal frequency of a Schwarzschild black hole in GR with no corrections from the higher--dimensional operators.
There are then two sources of corrections to account for.
First as derived in Section~\ref{sec:EFTBackground} the background black hole solution differs from that of GR. In particular in the EFT of gravity, the background black hole solution carries a horizon at $r_H$ rather than $r_g$. We shall thus denote by $\omega_{\rm GR}$ the quasinormal frequency of a black hole with Schwarzschild radius $r_H$, where $r_H$ relates to $r_g$ through Eq.~\eqref{eq:rH}. The second effect is in the corrections to the master equation as derived in Section~\ref{sec:masterEqs}.\\

\begin{figure}[th]
\centering
\includegraphics[width=0.7\textwidth]{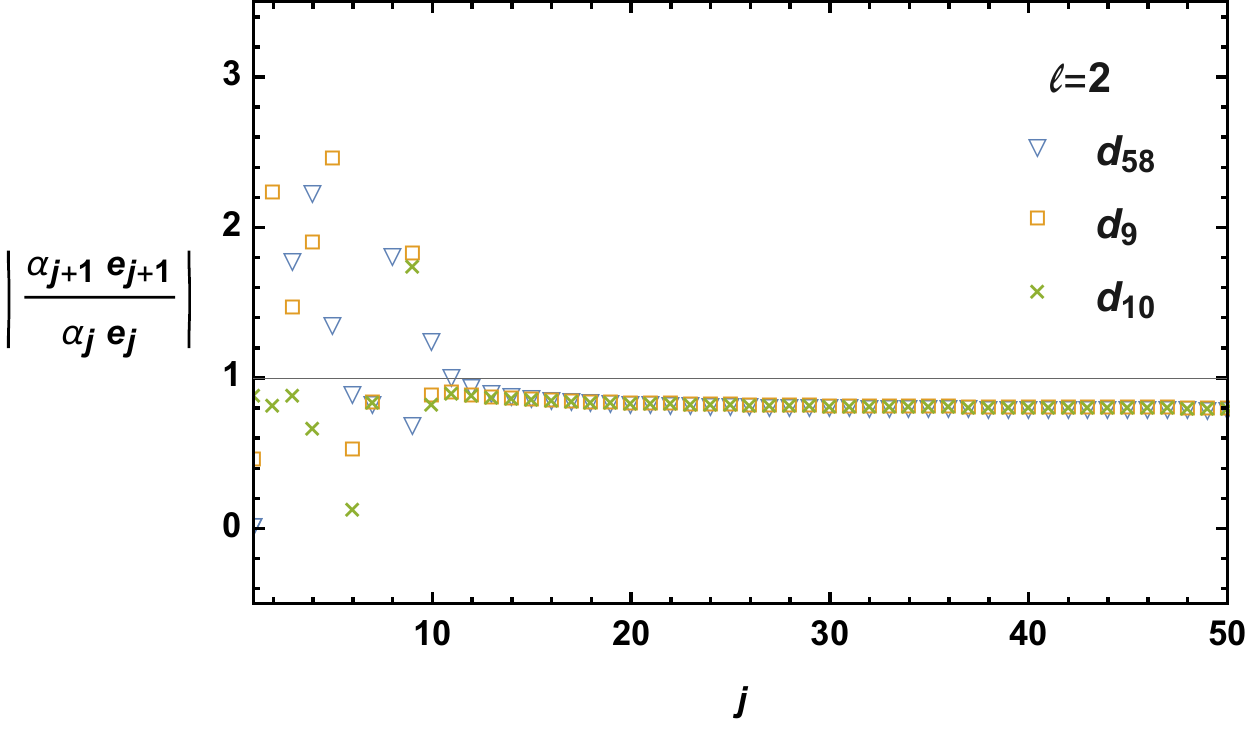}
\caption{The convergence of the corrections on the even quasinormal frequency for $\ell = 2$. The three different marks show the convergence of contributions from the $d_{58}$, $d_9$, and $d_{10}$ terms. As one can see in the plot, the ratio approaches $0.77$ as $j$ increases, which indicates the convergence condition \eqref{eq:converge} is satisfied. Also, the asymptotical ratio indicates that the fractional systematic error caused by the truncating on $j$ is ${\cal O}(10^{-3})$.} \label{fig:conv}
\end{figure}

Given the master equation for the odd and even tensor modes on the black hole background, Eq.~\eqref{eq:master}, computing the quasinormal frequency is then a straightforward procedure and one may follow any of the  many methods developed in the literature, see for example Ref.~\cite{Berti:2009kk} for a review. In this Section, we shall follow the method developed in \cite{Leung:1999rh,Leung:1999iq,Cardoso:2019mqo,Silva:2019scu}, and compute the leading corrections on the quasinormal frequency caused by the higher--dimension operators. The idea of this method is to make use of the asymptotical flatness of the background solution to parametrize the EFT corrections that enter the master equations \eqref{eq:GRmaster} as a power--law expansion of the form
\ba
V_{\rm para}^{\rm o/e} = V_{\rm GR}^{\rm o/e} + \delta V^{\rm o/e}, \quad {\rm with} \quad \delta V^{\rm o/e} = \frac{1}{r_H^2}\sum_{j=0}^{\infty} \alpha_j^{\rm o/e} \left(\frac{r_H}{r}\right)^j\,.
\ea
We will neglect the superscription ``o/e" in the following. At linear order, each term in $\delta V$ contributes to the quasinormal frequency independently, and the corrected quasinormal frequency can be written as
\ba
\omega = \omega_{\rm GR} + \sum_{j=0}^{\infty} \alpha_j  e_j,
\ea
where $e_j$ are complex numbers and have been calculated for $\ell \le 10$ and up to $j = 50$ in \cite{Cardoso:2019mqo}. As discussed in \cite{Cardoso:2019mqo}, the correction on the quasinormal frequency converges when
\ba\label{eq:converge}
\lim_{j\rightarrow \infty} \left| \frac{\alpha_{j+1}e_{j+1}}{\alpha_j e_j}\right| < 1.
\ea
In order to apply this method to our case, we first introduce the normalized variable $\tilde{\Psi} = \sqrt{N(r)} \, \Psi$ defined so that Eq. \eqref{eq:master} can be written as
\ba\label{eq:QNMeq}
f \frac{\rm d}{{\rm d} r} \left(f\frac{{\rm d \tilde{\Psi}}}{{\rm d} r}\right) + \bigg[\tilde{\omega}^2 - f V_{\rm para}\bigg] \tilde{\Psi} = 0\,,
\ea
with $f= 1-r_H/r$ and $\tilde{\omega} = \omega / \gamma$ being a rescaled frequency. Here we have defined
\ba
\gamma \equiv \left. c_s N \right|_{r =r_H} = 1 - \epsilon \left(3 d_{58}+6 d_9 - \frac32 d_{10}\right)  + {\cal O}(\epsilon^2).
\ea
The expression for $N(r)$ is given explicitly in Appendix~\ref{app:exp}. From Eq.~\eqref{eq:QNMeq}, we identify the corrections to the GR potential $\delta V = V_{\rm para} - V_{\rm GR}$, where $V_{\rm GR}$ is the GR potential of a Schwarzschild black hole with horizon located not at $r_g$ but rather at $r_H$. Recall that $r_H$ relates to $r_g$ through Eq.~\eqref{eq:rH}. $\delta V$ is a function of $r$, $r_H$, $J$ and $\tilde{\omega}$ in general. Since $\delta V$ is already a first order correction, we can replace $\tilde{\omega}$ with $\omega_{\rm GR}$ (or equivalently with $\omega_0$) within $\delta V$, where we recall that $\omega_{\rm GR}$ denotes the quasinormal frequency of a black hole with Schwarzschild radius $r_H$.  We can then read off $\alpha_j$ by expanding $\delta V$ as a Taylor series in $r_H/r$. For odd perturbations, we find $\delta V$ is a polynomial in $r_H/r$ with finite terms. For even perturbations, the convergence condition \eqref{eq:converge} is expected to be satisfied as shown in Fig.~\ref{fig:conv}. Therefore, the leading correction on the quasinormal frequency is given by
\ba\label{eq:domega}
\delta \omega \equiv \omega - \omega_0 =  \left(\gamma\frac{r_g}{r_H}-1\right) \omega_0 + \sum_{j=0}^{\infty} \alpha_j  e_j ,
\ea
where $\omega_0$ is the quasinormal frequency of a black hole with Schwarzschild radius $r_g$, i.e., the black hole with no correction from higher--dimension operators. In the last line of Eq.\eqref{eq:domega}, we have made use of relation $\omega_0 r_g  =  \omega_{\rm GR}r_H$. We calculate $\delta \omega$ by summing $\alpha_je_j$ up to $j = 50$, and present the result in Table~\ref{tab:corr} in the form of fractional corrections
\ba\label{eq:delta}
\delta \equiv  \left(\frac{\rm{Re}(\omega - \omega_0)}{\epsilon \, \rm{Re}(\omega_0)}, \, \frac{\rm{Im}(\omega - \omega_0)}{ \epsilon\, \rm{Im}(\omega_0)}\right) = d_{58} \delta_{58} + d_{9} \delta_{9} + d_{10} \delta_{10}.
\ea
In particular, we find the contribution from the term $d_{58}$ to be negligible compared to that of the genuine Weyl--cubed terms $d_9$ and $d_{10}$. Indeed, as shown in Table~\ref{tab:corr}, the value of  $|\delta_{58}|$ computed numerically is at least four orders of magnitude smaller than those for $|\delta_{9}|$ and $|\delta_{10}|$. Actually, we expect the contribution from $d_{58}$ to vanish entirely, i.e. we would expect $\delta_{58}\equiv 0$ exactly. As explained in Appendix~\ref{app:redef} when $d_9 = d_{10} = 0$, the theory~\eqref{eq:Ltotal}  is equivalent to GR in the vacuum up to ${\cal O}(1/M^4)$ corrections and the operator $d_{58}$ should therefore not contribute to any physical observable in the vacuum. The quasinormal frequency should therefore be insensitive to the coefficient $d_{58}$. For the odd modes, the power--law expansion of $\delta V^{\rm o}$ only includes a finite number of terms, and the non--vanishing value computed for $|\delta_{58}^{\rm o}r_g \omega_0| \sim {\cal O} (10^{-6})$ comes only from the numerical errors in $\omega_0$ and $e_j$, which is precisely of order $10^{-6}$. For the even modes, the numerical error of $|\delta_{58}^{\rm e}r_g \omega_0|$ is dominated by the truncation of the sum over $j$. For example, for $\ell = 2$, we see from Fig.~\ref{fig:conv} that at $j\sim 50$, we have $\rho =|\alpha_{j+1}e_{j+1}/\alpha_j e_j| \approx 0.77$. This suggests that truncating the sum over $j$ at $j=50$ leads to an error on the computed values of $\delta_{i}$  of order $|\alpha_{50}e_{50}| \rho/(1-\rho) \simeq 8.8\times 10^{-4} \sim {\cal O}(10^{-3})$. In other words, the small numerical value we obtain for  $|\delta_{58}^{\rm e}\omega_0r_g| < {\cal O}(10^{-3})$ is consistent with it vanishing within our error bars and actually justifies the convergence of our results. This represents a non--trivial check.

\begin{table}[tbp]
\centering
\begin{tabular}{lccc}
\hline
       	   &  $\ell = 2$ & $\ell = 3$ & $\ell = 4$   \\
\hline
\hline
$\delta^{\rm o/e}_{58}$ (theoretical)  &$(0, \,0)$  & $(0, \, 0)$   & $(0, \,0)$  \\
$\delta^{\rm o}_{58}\times10^6$  &$(1.22, \,-0.24)$  & $(-0.76, \, -0.87)$   & $(-0.26, \,-2.27)$  \\
$\delta^{\rm e}_{58}\times10^4$  &$(1.55, \,-0.84)$  & $(-0.04, \,0.06)$   & $(-0.02, \,-0.03)$  \\
$\delta^{\rm o}_{9}$  & $(21.07, \,47.53)$  & $(18.29, \,46.27)$ & $(17.54, \,45.71)$  \\
$\delta^{\rm e}_{9}$  & $(-12.33, \,-58.12)$  & $(-14.15, \,-51.82)$ & $(-14.46, \,-50.17)$  \\
$\delta^{\rm o}_{10}$ & $(10.54, \,23.76)$  & $(9.14, \,23.13)$ & $(8.77, \,22.85)$  \\
$\delta^{\rm e}_{10}$ & $(-6.17, \,-29.06)$  & $(-7.07, \,-25.91)$ & $(-7.23, \,-25.08)$  \\
\hline
\end{tabular}
\caption{Fractional corrections on the quasinormal frequency for the modes $\ell = 2$, $3$, and $4$. Corrections are computed using the method developed in \cite{Cardoso:2019mqo}, with a summation of $e_j$ up to $j=50$. $\delta$ is defined in Eq.~\eqref{eq:delta} with a superscription ``o" or ``e" that denotes corrections to the odd or even quasinormal frequency. The uncorrected frequencies are $r_g \omega_0 = 0.747343 -  0.17925\,i$, $1.199887-  0.185406\,i$, and $1.618357-0.188328\,i$ for $\ell = 2$, $3$, and $4$ respectively. Theoretically, we expect $\delta_{58}^{\rm o/e} = 0$. Upon explicit numerical computation we find a nonzero but negligible $|\delta_{58}^{\rm o/e}|$ which arises from numerical errors and from truncating the series expansion at $j=50$. For example, for $\ell = 2$, we have $|\delta^{\rm o}_{58} \omega_0 r_g | \sim {\cal O} (10^{-6})$, reflecting the numerical errors of $\omega_0$. As for the even mode, we have $|\delta^{\rm e}_{58} \omega_0 r_g | \sim {\cal O} (10^{-4})$, which is the order of magnitude we expect from truncating the sum over $j$ at $50$. The smallness of $|\delta_{58}^{\rm o/e}|$ justifies the convergence of our results. \label{tab:corr} }
\end{table}

\section{Discussion and Outlook}\label{sec:dis}

It is known that the speed of GWs could be different from the speed of photons due to interactions with other fields which may manifest themselves as irrelevant operators in the low--energy EFT of gravity (or of electromagnetism). In this paper, we investigate the propagation of GWs on a black hole background. We work with an EFT of gravity at low--energy, in which effects from other fields are captured by higher-dimensional curvature operators. We study perturbations around black holes in this theory, and derive the modified Regge--Wheeler--Zerilli equations~\eqref{eq:master}. We find the leading modification on the speed of GWs, which is entirely determined by only two of the dimension--6 operators, namely the pure Riemann or Weyl--cubed operators $d_9$ and $d_{10}$ in Lagrangian~\eqref{eq:l3}. We also compute the leading corrections of higher dimension operators on the quasinormal frequency, which are shown in Table~\ref{tab:corr}. Again, as expected, only the pure Weyl--cubed operators ($d_9$ and $d_{10}$) affect the quasinormal frequency as the other terms can be gauged away by a field redefinition. In particular, corrections from the $d_5$ and $d_8$ terms vanish in spite of their appearance in the perturbation equations~\eqref{eq:master}, and the effects from the parameters $d_9$ and $d_{10}$  are governed by the quantity $2 d_{9} + d_{10}$.\\

We find that the low--energy speed of GWs can be either superluminal or subluminal on a Schwarzschild--like black hole, depending on the precise coefficients of the Weyl--cubed operators that enter the low--energy EFT. These coefficients depend on the details of the heavy fields and more specifically on their precise spin. We show that at low--energy, GWs would see a different local causal structure as compared to photons or other minimally coupled species. Nevertheless, the departure vanishes at the black hole horizon, implying that the location of the horizon is identically defined for both GWs and photons.\\

Due to the hierarchy in scales, effects caused by the higher--dimension operators are suppressed by $\epsilon$ defined in Eq.~\eqref{eq:epsilon}. This leads to a typical suppression of order
\ba
\epsilon \sim 10^{-162} \left(\frac{M}{M_{\rm Pl}}\right)^{-2} \left(\frac{M_{\rm BH}}{M_\odot}\right)^{-4}\,,
\ea
where $M$ is the cut--off scale of the EFT, and $M_{\rm BH}$ the black hole mass. Even in the highly extravagant scenario where  we consider the higher--dimension operators to represent the loop corrections from say a  dark energy field, with $M$ as low as the Hubble scale, i.e., $M \sim 10^{-60} M_{\rm Pl}$, the corrections on astrophysical black holes would still be incredibly suppressed. For instance even in that scenario we would get $\epsilon \sim  10^{-48}$ for $M_{\rm BH} = 30 M_\odot$. In this sense, the purpose of this work is not to predict observable effects that can be tested with astrophysical black holes, but to understand the effects from operators that naturally enter the EFT of gravity on the causal structure of GWs from a theoretical point of view. Nevertheless, the effects from the higher dimension operators could  be more significant for very small black holes. For example, we would have $\epsilon \simeq 0.1$ for $M_{\rm BH} \simeq 8 \times 10^{-11} M_\odot \sim  10^{23} {\rm g}$. Such light black holes could form in the early universe, known as primordial black holes. Even though such light black holes  have not been observed, primordial black holes are constrained by their possible observational effects \cite{Carr:2020gox}. \\

In this study, we only investigate the leading corrections and observe that the horizon as seen by low--energy GWs remains the same as the higher frequency ones and as that of  other massless particles.  This appears to be no accident and one would expect the same to remain valid to all order in the EFT expansion although this has not been checked explicitly in this study. At higher order in the EFT expansion, we   expect to observe a  frequency dependence in the modified dispersion relation and leading to a speed of GWs to return to exact luminality at large frequencies. Although this is beyond the scope of this work, it  would be interesting to check this behaviour explicitly.\\

If the EFT of gravity is obtained from integrating out heavy fields of spin smaller than 2, it can be shown that wether or not GWs are effectively superluminal (with respect to the vacuum speed of light) depends on the spin of the lightest massive particle that has been integrated out. In the case of fermions, we see that the effective speed of GWs outside the horizon is always superluminal. This type of argument has been used in the past to discriminate against various types of EFTs. Applied to the present context this would naively suggest that fermions cannot be the lightest massive particles and therefore there must exist another massive field (of spin other than $1/2$) with mass very close to that of neutrinos (it cannot be arbitrary low otherwise the EFT of gravity at energy scales between the mass of this new particle and that of the neutrinos would suffer from the same issue). This appears to be a remarkably strong conclusion and we would warn against this type of arguments. Instead we emphasize that this effective low--energy superluminality is not always connected to microscopic violation of causality \cite{Hollowood:2015elj,Hollowood:2007kt,Hollowood:2007ku,Hollowood:2008kq,Hollowood:2009qz,Hollowood:2010bd,Hollowood:2010xh,Hollowood:2011yh,Hollowood:2012as,Goon:2016une,deRham:2020zyh}.

\acknowledgments

We would like to thank Pablo A. Cano, Vitor Cardoso, Masashi Kimura, Andrea Maselli and  Andrew J. Tolley for the useful discussions. The work of CdR is supported by an STFC grant ST/P000762/1. CdR thanks the Royal Society for support at ICL through a Wolfson Research Merit Award. CdR and JZ are supported by the European Union's Horizon 2020 Research Council grant 724659 MassiveCosmo ERC-2016-COG. CdR is also supported by a Simons Foundation award ID 555326 under the Simons Foundation's Origins of the Universe initiative, `\textit{Cosmology Beyond Einstein's Theory}'. JF is supported by the Fonds National Suisse. JF thanks as well the support of Fondation Boninchi. JF would like to thank Imperial College London for its hospitality during this work.

\appendix

\section{Relevant Operators and Field Redefinition}\label{app:redef}

The EFT of gravity includes various dimension--4 and --6 operators (as well as of course an infinite number of other higher dimension operators). In principle all of those contribute in non--trivial way to the modified background solution and to the dynamics of GWs, however as already emphasized in Section~\ref{sec:BH}, for the vacuum solution we are interested in, most of these operators are irrelevant (in the sense that they do not contribute either to the modified background solution or to the evolution of GWs).\\

First expressed in terms of the Ricci scalar and tensor as in \eqref{eq:D4} it is clear that none of the dimension--4 operators can contribute to the Ricci--flat solution. Moreover,  not all of the dimension--6 curvature operators contribute at leading order. Actually, any term in the Lagrangian that is of second power of the Ricci tensor does not contribute to the Ricci flat solution at leading order (i.e. at first order in $\epsilon$). This is because the equations of motion generated by such terms is proportional to $\epsilon R_{\mu\nu}$ which vanishes at first order in $\epsilon$ if the background solution is Ricci flat at zeroth order in $\epsilon$.

Moreover, for these particular types of solutions, the operators $d_5$ and $d_8$ are not independent and always appear as the combination $d_{58} \equiv 4d_5 + d_8$. This can be understood  by rewriting the $d_5$ and $d_8$ terms in \eqref{eq:l3} as
\ba\label{eq:L58}
\mathcal{L}_{5,\,8} = \frac{1}{M^2}\sqrt{-g} \left[
d_5 (g_{\mu\nu} R_{\mu \nu \alpha\beta}^2  - 4 R_{\mu\alpha\beta\gamma}\tensor[]{R}{_\nu^{\alpha\beta\gamma}})R^{\mu\nu} +  d_{58} R^{\mu\nu}R_{\mu\alpha\beta\gamma}\tensor[]{R}{_\nu^{\alpha\beta\gamma}} \right].
\ea
One can check that $g_{\mu\nu} R_{\mu \nu \alpha\beta}^2  - 4 R_{\mu\alpha\beta\gamma}\tensor[]{R}{_\nu^{\alpha\beta\gamma}}$ vanishes for the Schwarzschild metric, therefore varying $\mathcal{L}_{58}$ with respect to $g^{\mu\nu}$, one can find that the first term on the right hand side of Eq. \eqref{eq:L58} does not contribute to the equation of motion at the first order in $\epsilon$.\\

We further note that there exists a change of frame that allows us to field--redefine most of the dimension--6 operators introduced in the EFT \eqref{eq:Ltotal}. Note however that after field redefinitions, these operators appear as non--minimal coupling to all matter fields, including the photon and other massless particles. The following perturbative field redefinition of the metric
\ba
g_{\mu\nu} \rightarrow g_{\mu\nu} - \frac{2}{\mpl^2} \delta g_{\mu\nu}\,,
\ea
modifies the Einstein--Hilbert action by
\ba
\delta \mathcal{L}_{\rm EH} =\sqrt{-g} R^{\mu\nu} \left(\delta g_{\mu\nu} - \frac12 g_{\mu\nu} \delta g\right)\,.
\ea
We can perform a field redefinition
\ba
g_{\mu\nu} &\rightarrow& g_{\mu\nu} - \frac{2}{\mpl^2} \left[-2 c_{W^2} R_{\mu\nu} + \left(c_{R^2}+\frac13 c_{W^2}\right)g_{\mu\nu}R\right] \\
&-& \frac{2}{\mpl^2}\frac{1}{M^2}\bigg\{ - d_2 \Box R_{\mu\nu} -d_4 R R_{\mu\nu}  -d_6 {R_{\mu}}^{\alpha}R_{\nu\alpha}   -d_7  R_{\alpha\beta}R_{\mu\alpha\nu\beta} -d_8 {R_{\mu}}^{\alpha\beta\gamma}R_{\nu\alpha\beta\gamma} \nonumber  \\
 &+& g_{\mu\nu}\left[\left(d_1+\frac{d_2}{2}\right) \Box R + \left(d_3+\frac{d_4}{2}\right)R^2 + \left(\frac{d_6}{2}+\frac{d_7}{2} \right) R_{\alpha\beta}^2 + \left(d_5+\frac{d_8}{2}\right)R_{\alpha\beta\gamma\sigma}^2   \right] \bigg\},
 \label{eq:Fieldg}
\ea
so that only the operators that are genuinely Weyl--cubed (or Riemann--cubed) are left in Lagrangian \eqref{eq:Ltotal}, i.e.,
\ba\label{eq:Lnew}
\mathcal{L} &=& \sqrt{-g} \left[ \frac{\mpl^2}{2} R +  \frac{1}{M^2} \left( d_9 \tensor{R}{_\mu_\nu^\alpha^\beta} \tensor{R}{_\alpha_\beta^\gamma^\sigma} \tensor{R}{_\gamma_\sigma^\mu^\nu}+ d_{10}\tensor{R}{_\mu^\alpha_\nu^\beta}\tensor{R}{_\alpha^\gamma_\beta^\sigma} \tensor{R}{_\gamma^\mu_\sigma^\nu} \right) \right] +{\cal O}\left(\frac{1}{M^4} \right) .
\ea
Although such a change of frame simplifies the Lagrangian, this field redefinition introduces non--minimal couplings between gravity and matter fields \cite{deRham:2019ctd} which in turn affect the effective metric seen by matter fields. Rather than needing to account for those minimal couplings, in  the main part of this work we find it more convenient to work directly    in the original frame where all the operators \eqref{eq:l3} appear and the speed of photons is unity.\\

In what follows, we provide a preview of the analysis in the new frame with Lagrangian \eqref{eq:Lnew}, before returning to the original frame by a reversal redefinition. This is particularly convenient when $d_9 = d_{10} = 0$. In that special case, the new frame would be equivalent to GR with no other corrections at this order in the EFT in the absence of other matter fields. It directly follows that any frame independent quantities such as the quasinormal frequency should be the same as GR up to ${\cal O}(1/M^4)$ corrections. In particular, we can conclude that aside from $d_9$ and $d_{10}$ no other dimension--6 (and --4) operators should affect the quasinormal frequency at linear order in $1/M^2$.

However in the general case where $d_9 \neq 0$ or $d_{10} \neq 0$, the  black hole solution in the new frame is obtained in a different coordinate system with radial coordinate $\tilde r$, which relates to original one through
\ba
\tilde{r} = r \left(1 - \epsilon 3 d_{58} \frac{r_g^6}{r^6}\right).
\ea
This explains why the propagation of GWs (whose speed is not frame independent\footnote{To be more precise, the low--energy speed of GWs is not frame independent but we expect the ratio of the low--energy speed of GWs to that of photons to be frame independent. Note however that in the frame where the Lagrangian for gravity takes the form \eqref{eq:Lnew}, photons are no longer minimally coupled. In that frame, the low--energy speed of GWs will only depend on $d_9$ and $d_{10}$, but the low--energy speed of photons is expected to  depend on the coefficients $d_5$ and $d_8$ (the coefficients governing the pure--Weyl terms in the field redefinition \eqref{eq:Fieldg}). It should therefore come as no surprise that even in that frame the ratio of the speed of GWs and photons depends not only on $d_9$ and $d_{10}$ but also on $d_5$ and $d_8$.}) ends up depending not only on the coefficients of the pure Weyl--cubed terms $d_9$ and $d_{10}$ but also on the contributions from the operators $d_{5}$ and $d_8$. Accounting for these contributions can be done either via an explicit inverse change of coordinate or by working directly with all the operators present in the EFT from the outset.

Moreover, in four dimensions, the operator $d_{10}$ can be written as a combination of other dimension--6 operators (see the appendix in Ref.~\cite{Cano:2019ore} for more details). As one can check in our results, only the coefficients $2d_{58} - 3 d_{10} $ and $2d_9+ d_{10}$ contribute to the background solution and the dynamics of GWs.

\section{Explicit Expressions}\label{app:exp}

In this appendix, we give some of the key steps that underwent the derivation of Eq.~\eqref{eq:master} as well as the explicit expressions for some of the relevant functions used in the master equation.

Since within the regime of validity of the EFT, the higher--dimensional curvature operators ought to be treated perturbatively, one can simply start by using the same Einstein equations that lead to the Regge--Wheeler--Zerilli equations, which yield two master equations (one for the odd perturbations, and one for the even perturbations) in terms of $h_0$, $h_1$, $H_0$, $H_1$, $H_2$ and ${\cal K}$.
Including the leading corrections from the EFT higher--order operators then leads to  extra terms in our master equations that are linear in $\epsilon$. Any higher derivatives proportional to $\epsilon$ can (and indeed {\it should} in this EFT) be removed using the lower order GR perturbation equations. In order to rewrite our master equations in the form of Eqs.~\eqref{eq:master}, we define our master variables as Eqs.~\eqref{eq:variable}, allowing deviations from the GR definitions at ${\cal O}(\epsilon)$. The derivations are described by $f_{h_1}$, $f_\K$ and $f_{H_1}$ with explicit forms to be determined below. In the subsections \ref{subapp:odd} and \ref{subapp:even} we provide further details on how to determine these functions for the odd and even perturbations.

Substituting Eqs.~\eqref{eq:variable} into Eqs.~\eqref{eq:master} yields an odd and even master equation which similar to that in GR with the addition of ${\cal O}(\epsilon)$ corrections.  The precise expressions of the correction to the effective potential $V^{\rm o/e}$ as well as $f_{h_1}$, $f_\K$ and $f_{H_1}$ can be determined by matching the coefficients of $\K$, $H_1$ and their derivatives to the master equations obtained previously.

\subsection{Odd perturbations}
\label{subapp:odd}

Making use of the Regge--Wheeler gauge \cite{Regge:1957td}, the odd perturbations  involve two variables $h_{0,1}$ introduced in \eqref{eq:habodd}. In GR, Einstein's equations include one constraint that can be used to identify  $h_0$ and its first derivatives  in terms of $h_1$ and its first derivatives. The only remaining independent Einstein's equation for the odd perturbations then provides the   master equation as a second order evolution equation for  $h_1$. In the EFT at hand, we proceed in a similar way. As mentioned previously, we make use of the GR equations to remove any higher order derivatives.  The master equation then takes the similar form as in GR, namely involving only $h_1$ and at most its second derivatives.
At this point, one can perform a field redefinition of the form
\ba
\Psi^{\rm o} = \frac{i \sqrt{A B}\, h_1}{r \omega} \bigg[1+ \epsilon f_{h_1} \bigg]\,,
\ea
and the expression for $f_{h_1}$ can be found by requesting the master equation to be written in the following form,
\ba
\sqrt{AB} \frac{{\rm d}}{{\rm d}r}\left(\sqrt{AB}\, \frac{{\rm d} \Psi^{\rm o}}{{\rm d} r}\right) + \omega^2 \Psi^{\rm o}(r) - \sqrt{AB}\, \left[V_{\rm GR}^{\text{o}} + \epsilon\,  {\cal V}^{\rm o}(r, \omega) \right] \Psi^{\rm o}(r) =0.
\ea

\subsection{Even perturbations}
\label{subapp:even}
Obtaining the master equation for the even mode needs more care. Making use of the Regge--Wheeler gauge \cite{Regge:1957td}, the even perturbations now  involve four variables namely the quantities $H_{0,1,2}$ and $\K$ introduced in \eqref{eq:evenhab}. Using the same Einstein equations that led to the Regge--Wheeler--Zerilli equations, one gets an equation of the form
\ba
{\cal E} = {\cal E}^{\rm GR}+  \epsilon\, \delta{\cal E} = 0,
\ea
where ${\cal E}^{\rm GR}$ is the GR master equation in terms of $H_1$, $\K$ and their derivatives, and $\delta{\cal E}$ is a function of $H_{0,1,2}$, $\K$ and their derivatives. $H_{0,2}$ in $\delta{\cal E}$ can be replaced with $H_{1}$ and $\K$ using the GR perturbation equations, after which we have ${\cal E}$ being an equation of only $H_1$ and $\K$.

On the other hand, we can define the master variable as
\ba
\Psi^{\rm e} = \frac{1}{(J-2)r + 3 r_g}\left[-r^2 \K (1+\epsilon f_\K)+ \frac{i \sqrt{AB}\,rH_1}{ \omega}(1+\epsilon f_{H_1})\right]\,,
\ea
with unknown functions $f_{H_1, \K}$. Substituting it into Eq.\eqref{eq:master} gives
\ba
\sqrt{AB} \frac{{\rm d}}{{\rm d}r}\left(\sqrt{AB}\, \frac{{\rm d} \Psi^{\rm e}}{{\rm d} r}\right) + \frac{\omega^2}{c_s^2} \Psi^{\rm e}(r) - \sqrt{AB}\, \left[V_{\rm GR}^{\text{e}} + \epsilon\,  V^{\rm e}(r) \right] \Psi^{\rm e}(r) = {\cal E} + \epsilon \sum f_i (r) {\cal E}^{\rm GR}_i =0,
\ea
where $\sum f_i (r) {\cal E}^{\rm GR}_i$ denotes a linear combination of GR perturbation equations. Then the unknown functions $f_{H_1, \K}$, $V^{\rm e}$ and the linear combination $f_i$ can be determined by identifying the coefficient of $\K$, $H_1$ and their derivatives on both sides of the equation.

\subsection{Explicit expressions}

Following the prescriptions described above, we find  after explicit calculations the  $V^{\rm o/e}$, $f_{h_1}$, $f_\K$ and $f_{H_1}$ defined in Eqs.~\eqref{eq:master} given by
\ba
 V^{\rm o} &=& \frac{3}{x^9 r_g^2} \bigg \{ d_{58} (24 - 7 J x)  +  6 d_9  \bigg[80 (J-6) x^2+5 (230-19 J) x - 662 \bigg]   \\
&&+ \frac{3 d_{10}}{2}  \bigg[160 (J-6) x^2+(2300-183 J) x -1348\bigg] \bigg \}, \nonumber
\\
V^{\rm e} &=& \frac{-3}{2 x^9 [(J-2) x+3]^3r_g^2}  \\
&&\bigg \{d_{58}\bigg[14 (J-2)^3 J x^4+6 (J-2)^2 (13 J-16) x^3+270 (J-2)^2 x^2+558 (J-2) x+432\bigg] \nonumber\\
&&+d_9\bigg[-36 (J-2)^2 \left(15 J^2-336 J+836\right) x^4-60 (J-2) \left(147 J^2-1304 J+2164\right) x^3 \nonumber \\
&&+480 (J-6) (J-2)^3 x^5-36 (J-2) (1073 J-3988) x^2-12 (5407 J-13460) x-36648\bigg] \nonumber \\
&&+ d_{10} \bigg[-3 (J-2)^2 \left(97 J^2-2030 J+5016\right) x^4-3 (J-2) \left(1509 J^2-13166 J+21736\right) x^3\nonumber \\
&&+240 (J-6) (J-2)^3 x^5-9 (J-2) (2191 J-8066) x^2-3 (11093 J-27478) x-18972\bigg]\bigg\}, \nonumber \\
f_{h_1} &=&  \frac{3}{2x^6}\bigg[2 d_{58} + d_9 \left(64-96x\right) + d_{10} \left(29-48x\right)\bigg], \\
f_{\K}\, &=& \frac{1}{x^6\left[(J-2)x + 3\right]} \\
&&\bigg \{ 3 d_{58} \bigg [2 (J-2) J x^2+(J+10) x-12 \bigg] + 6 d_9 \bigg[24 (J-2) x^2-4 (8 J-25) x-63\bigg] \nonumber \\
&&-\frac{3}{2}  d_{10} \bigg [6 \left(J^2-10 J+16\right) x^2+(67 J-170) x+90\bigg] \bigg\}, \nonumber
\ea
\ba
f_{H_1} &=& \frac{1}{x^6\left[(J-2)x + 3\right]} \\
&&\bigg \{ 3 d_{58} \bigg [2 (J-2) J x^2+(8 J-4) x+9 \bigg] + 6 d_9 \bigg[24 (J-2) x^2+(82-23 J) x-36\bigg] \nonumber \\
&&-\frac{3}{2}  d_{10} \bigg [6 \left(J^2-10 J+16\right) x^2+2 (35 J-88) x+99\bigg] \bigg\}, \nonumber
\ea
where $x \equiv r/r_g$. The field redefinition $\tilde{\Psi} = \sqrt{N(r)}\Psi$ defined above Eq.~\eqref{eq:QNMeq} is given by
\ba
N(r) = 1 - \frac{\epsilon}{x-1} \left[3d_{58} \left(\frac{4}{x^6}-\frac{5 }{x^5}+1\right)+d_9 \left(\frac{44}{x^6}-\frac{54}{x^5}+10\right)+d_{10} \left(\frac{4}{x^6}-\frac{9}{2 x^5}+\frac{1}{2}\right)\right].
\ea

\bibliographystyle{JHEP}
\bibliography{master}
\end{document}